\documentclass[doublecol]{epl2} 

\usepackage{graphicx}
\usepackage{array}
\usepackage{amssymb}
\usepackage{dcolumn}
\usepackage{bm}
\usepackage{color}

\def\etal{{et al}. }

\newcommand{\lbar}{\left|}
\newcommand{\rbar}{\right|}

\title{Games on graphs: A minor modification of payoff scheme makes a big difference}
\shorttitle{Games on graphs: A minor modification makes a big difference} 

\author{Qiang Zhang \and Tianxiao Qi \and Keqiang Li \and Zengru Di \and Jinshan Wu $^{\dag}$}
\shortauthor{Q. Zhang \etal}

\institute{                    
  School of Systems Science, Beijing Normal University, Beijing, P.R. China, 100875
}

\abstract{
Many techniques developed in simulations of physical models have been adopted in studies of game theory by researchers including physicists and mathematicians. In this work, we show that a seemly non-essential mechanism --  what we refer to as a ``payoff scheme'' have a large impact on strategic outcomes of some games. Payoff scheme refers to here that how each player's payoff is calculated in each round after the states of all of the players are determined. Conventionally either the accumulated or the average payoff of a player is used, where its payoff is calculated from pairing up the player with all of its neighboring players. Here we consider to calculate the payoff from pairing up with only one random player from the neighboring players. The average payoff scheme that involves averaging over all of the neighbors should, in a sense, be equivalent to repeatedly randomly pairing up with one neighbor a time, which we refer to as the stochastic payoff scheme. However, our simulation of games on graphs shows that, in many cases, the two payoff schemes lead to qualitatively different levels of cooperation: Seemly non-essential modifications might have large impact on behavioral outcomes. We have also observed that results from the stochastic scheme are more robust than the average scheme: different updating rules and initial states of the players do not have a large impact on the final level of cooperation in the former case when compared with those in the latter case. 
}

\begin{document}

\maketitle

\section{Introduction}
The emergence of cooperation has been one of the central topics in game theory and its application in social studies, human behavior and biology \cite{sugden1986economics, sigmund1993games}. Understanding the relatively high level of cooperation among inherently selfish players remains a challenge, especially in situations in which there is a social dilemma, where the theories of game predict defection but not cooperation as the solution to the games ({\it{i.e.}}, the theoretically expected game outcome). However, it is well recognized that, in many social dilemmas, cooperation is observed much more frequently than what the theories predict. Many natural and social scientists were inspired to investigate possible mechanisms of the emergence of cooperation\cite{axelrod1981evolution, smith1993evolution}. Thus far, it turns out that evolutionary game theory \cite{weibull1997evolutionary, herbert2000game, nowak2006evolutionary} in well-mixed or heterogeneously localized (on lattices or networks) populations provides the most general theoretical framework for this line of investigation. 

In evolutionary game theory, symmetric $2\times2$ games, such as the Prisoner's Dilemma (PD) and the Snowdrift Game (SG), have been used comprehensively as the underlying social dilemma for studies of the evolution of cooperation. A symmetric $2 \times 2$ game can be represented by a payoff bi-matrix as
\begin{equation}
 G^{C,D}= \left[\begin{array}{cc}
   R,R & S,T \\
   T,S & P,P
   \end{array}\right],
\end{equation}
where each of the two players has two strategies, Cooperation ($C$) and Defection ($D$). Mutual cooperation yields the reward $R$, mutual defection leads to the punishment $P$, the mixed choice gives the cooperator the sucker's payoff  $S$, and the defector
has the temptation $T$. To simplify, the symmetric $2\times2$ game is usually rescaled such that $R>P$ , $R=1$ , and $P=0$. The different game situations are determined by different values of the parameters $S$ and $T$.

The well-known PD corresponds to the case of $T>R>P>S$, where evolutionary game theory of a well-mixed population predicts that mutual defection is a single stable equilibrium. According to evolutionary game theory (also Nash game theory), in this game, cooperative behaviors can only occur when there are some unexplored additional mechanisms, such as heterogeneously localized interaction among players. This approach has developed into so-called spatial selection by Nowak and May \cite{nowak1992evolutionary, nowak1993spatial} and others \cite{lindgren1994evolutionary, durrett1994importance, killingback1996spatial}. It has been shown by their work and many follow-up studies \cite{szabo1998evolutionary, lieberman2005evolutionary, du2008evolutionary, allen2012mutation} that cooperation can occur in a certain situation, depending on the following three major elements \cite{szabo2007evolutionary, perc2010coevolutionary}: the underlying $2\times 2$ games \cite{hauert2004spatial, doebeli2005models}, the updating rules of players' strategic states \cite{tomassini2006hawks, ohtsuki2006simple, wu2007evolutionary} and topological structures of the underlying interaction networks \cite{nowak1992evolutionary, tomassini2006hawks, abramson2001social, szolnoki2007cooperation}. A few investigations also mentioned the effects of payoff schemes \cite{szolnoki2008towards, luthi2009evolutionary}.

Here, payoff schemes refer to how a player's payoff is calculated after all of the players made strategic decisions. When a player interacts with more than one neighbor, the player's payoff can not be directly determined by the original one-shot payoff matrices $G^{C,D}$. Conventionally, there are two schemes: the accumulated payoff and the average payoff. In both cases, in each round each player interacts pair-wisely with all of its neighbors. Then, the accumulated payoff of a certain player is computed by summing the payoffs from all $2 \times 2$ games with each of its neighbors. The average payoff is given by dividing the accumulated payoff by the number of its neighbors. 

Szolnoki et al. \cite{szolnoki2008towards} studied the effects of a combination of the schemes of the accumulated payoffs and average payoffs. They introduced a probability parameter $\alpha$ to combine the two schemes, with the probability $\alpha$ varying from $0$ to $1$ to represent the transition from the accumulated payoffs to the average payoffs. For PD on scale-free networks, they showed that with an increasing value of $\alpha$, the fraction of cooperators, which is relatively high due to the heterogeneous nature of the nodes of scale-free networks, deteriorates continuously, eventually collapsing. Noticing the drawbacks of the accumulated payoff and the average payoff, being that the former allows a player with a large degree to be extremely active, while the latter levels out too much of the benefit of having a large degree, Luthi et al. \cite{luthi2009evolutionary} proposed a modified payoff scheme with a guaranteed minimum payoff. They showed that the modified payoff scheme still allows players with large degrees to have a considerable amount of benefit. 

Although those payoff schemes, including accumulated, average, combination and the scheme modified by the guaranteed minimum payoff, are all different from each other, all of them are calculated over all of one's neighbors at each round. Thus, a player with $k$ neighbors participates in $k$ games at each round. This arrangement not only causes the sets of the neighbors to be different for each player but also makes the chance to act, thus the degree of activeness, to be different for each player. 

We argue that the spatial topological structure specifies only who can interact with whom, {\it{i.e.}} the players who are ``reachable''. It does not necessarily imply that one player is required to interact with all of its neighbors at each round. Here, for simplicity of terminology, the latter is referred to as ``reached''. The number of reached players depends not only on the reachable players but also on the activeness of the player who is reaching out. Noting the difference between the reachable and reached players, here we propose a new payoff scheme, which we call the ``stochastic payoff scheme''. Every round, a reaching player selects randomly one neighbor player from all of its neighbors and interacts with it. The payoff of the reaching player is calculated according to the original payoff matrices of the $2 \times 2$ game $G^{C,D}$. The payoff of the reached player will have to be calculated when it becomes a reaching player. 

It is the major task of the current investigation to compare the effect of this stochastic payoff against other payoff schemes, here mainly the average scheme, on cooperative behaviors. We argue that this randomly pairing up with one of the reachable players at each round is more reasonable than the alternative of reaching to all reachable players. Our consideration is that the attention or activeness of each player is limited, and it is acceptable to assume that this limitation is approximately the same for every player regardless of whether it has a larger or smaller set of reachable players. We also would like to note that this setting does not require too much activeness of the players that have large degrees; at the same time, it still allows players with large degrees to have a reasonable amount of advantage in both choosing from and observing the status of a large number of players.

A similar distinction between the set of players to play with and the set of players in observing status has been noted in the literature. Ohtsuki et al. \cite{ohtsuki2007evolutionary, ohtsuki2007breaking} proposed the idea of breaking the symmetry between sets of players during the stage of determining payoffs and updating strategies. In a sense, what we are proposing is a way to break the symmetry, also: one of the neighboring players is chosen in the first stage,  and all of the neighboring players are useful resources in updating one's strategy. 

There is another motivation of this investigation. Quite often we see different results reported from games on graphs that have very similar settings \cite{szabo2007evolutionary, perc2010coevolutionary}. The results reported range from the whole spectrum of significant boosting, marginally effecting to seriously decreasing the level of cooperation. For example, Nowak and May \cite{nowak1992evolutionary, nowak1993spatial} found that for PD on a two-dimensional lattice, cooperation can emerge and persist stably, while Hauert and Doebeli \cite{hauert2004spatial} showed that for SG, the evolution of cooperation is often inhabited. These results tell us that spatial network structures do not necessarily facilitate the level of cooperation. Santos and Pacheco \cite{santos2005scale} presented that a scale-free network, because of its heterogeneous nature, makes cooperation become the dominating trait, while Wu et al. \cite{wu2007evolutionary} found that when using the average payoff rather than the accumulated payoff, the advantage of a scale-free network is dismissed. Fu et al. \cite{fu2007evolutionary} and Yang et al. \cite{han2008evolutionary} observed that the optimal cooperation level exists for some moderately heterogeneous cases, but not the most heterogeneous or the most homogeneous cases. Experimental results that are qualitatively different from theoretical prediction have also been reported. For example, Cassar \cite{cassar2007coordination} discovered that for PD, cooperation was difficult to reach on local, random and small-world networks. Gruji{\'c} et al. \cite{grujic2010social} found that cooperation was not promoted by the existence of a lattice in most cases. Similar results were also yielded by Kirchkamp and Nagel \cite{kirchkamp2007naive}. The latest experiments \cite{gracia2012heterogeneous} show further that, when humans play a PD, heterogeneous networks do not boost cooperation, and they imply that for human beings, the spatial network structure has little relevance to the cooperators' promotion or inhibition. In experiments, the payoff scheme might be an issue that has a certain degree of importance and, to the best of our knowledge, it has not been investigated extensively.

In all of those different reported results, there is certainly something different in the settings, but we do not know which of the settings make the key difference that is responsible for generating the difference and whether it is possible that something that appears to be truly insignificant makes the largest contribution to the different observations. Here, we choose to question the effects of the various payoff schemes: rather than allowing players to play games with all of their neighbors and to obtain average/accumulated payoffs, what if each player pairs up with only one other player from its reachable neighbors and receives payoff accordingly? On average, this modification of the settings should result in nothing essential because, on average, pairing randomly with one other player is very much like playing with all of the neighbors with equal probability, thus seems to be equivalent to the average payoff scheme. On the other hand, it might not be this simple given that there is neither an energy function of the whole system as Hamiltonians for physical systems nor a principle of detailed balance for game systems. In simulation of Ising model, it is the overall Hamiltonian and the detailed balance that guarantee different orders of updating spins lead to the same stationary states. As will be seen later, this apparent non-essential factor does have an important impact on the observed behaviors. This finding indicates that, in line with the work on games on graphs, minor differences in settings can result in qualitatively different observations and when working on numerical studies of games extra attentions should be payed due to the lack of an overall energy function and the principle of detailed balance. 

This study is organized as follows. In Section $2$, we describe the evolutionary game model as well as five strategic updating rules in detail. In Section $3$, simulation results are provided, and in Section $4$, we summarize our main observations and discuss their implications. 

\section{Model}

The models that are considered in this work are the usual games on graphs, in which we have an underlying network and one player on each node of the network; each player has a set of strategies to choose from, and then, after their strategic statuses are decided, the payoffs to all of the players are calculated. This procedure is an iterative/evolutionary process: during the next round, the players update their strategies according to certain rules, and then, the payoffs are calculated again. The game setup is based on general $2 \times 2$ games that are defined by the payoff matrices $G^{C,D}$. The networks that we considered here include the regular von Neumann-neighbourhood \cite{nowak1992evolutionary, szabo2007evolutionary, hauert2002effects} (Moore-neighborhood \cite{szabo2007evolutionary, hauert2002effects}) lattice with $4$ ($8$) nearest neighbors for each node(respectively), the Watts-Strogatz (WS) small-world network \cite{watts1998collective} and the Barabasi-Albert (BA) scale-free network \cite{albert2002statistical, barabasi1999emergence}. These four networks are denoted, respectively, as networks $n=0, 1, 2$, and $3$. The payoff scheme that we use is the stochastic payoff scheme (denoted as payoff scheme $p=0$). The results from this scheme will be compared against those from the average payoff scheme (which is denoted as the payoff scheme $p=1$). Such a comparison is performed on games that have various payoff matrices, on various networks and under various rules of updating strategies. Those updating rules include imitating the best \cite{nowak1992evolutionary, nowak1993spatial}, imitating the better with an exponential probability function \cite{szabo1998evolutionary, traulsen2007pairwise}, imitating the better with a linear probability \cite{hauert2002effects, szabo2007evolutionary}, proportional update with an exponential probability function \cite{szabo1998evolutionary, traulsen2007pairwise} and proportional update with a linear probability \cite{hauert2002effects}. These five updating rules are denoted, respectively, as $u=0, 1, 2, 3$, and $4$. 

For convenience, we summarize here all of those updating rules by using the transition rates $\omega(s_{i} \rightarrow s_{j})$ for which player $i$ adopts the strategy of neighbor $j$.
\begin{enumerate}
\item Imitate the best, 
\begin{equation}
\omega(s_{i} \rightarrow s_{j}) =  \left\{
  \begin{array}{cc}
   \frac{1}{d_{i}} & E_{j} = max\left\{E_{k}\left|k \in N_{i}\cup \left\{i\right\}\right.\right\}\\
   0       & others
  \end{array}, \right.
\end{equation}
where $d_{i}$ is the number of equal maximum payoffs.
\item Imitate the better with an exponential probability, 
\begin{equation}
\omega(s_{i} \rightarrow s_{j}) = \frac{e^{\beta\cdot max(E_{j}-E_{i},0)}}{\sum_{k \in N_{i}} e^{\beta \cdot max(E_{k}-E_{i},0)}}.
\end{equation}
\item Imitate the better with a linear probability,
\begin{equation}
\omega(s_{i} \rightarrow s_{j}) = \frac{max(E_{j}-E_{i},0)}{\sum_{k \in N_{i}} max(E_{k}-E_{i},0)}.
\end{equation}
\item Proportional update with an exponential probability,
\begin{equation}
\omega(s_{i} \rightarrow s_{j}) = \frac{e^{\beta\left(E_{j}-M\right)}}{\sum_{k \in N_{i} \cup \{ i \}} e^{\beta\left(E_{k}-M\right)}}.
\end{equation}
\item Proportional update with a linear probability,
\begin{equation}
\omega(s_{i} \rightarrow s_{j}) = \frac{E_{j}-M}{\sum_{k \in N_{i} \cup \{ i \}} (E_{k}-M)}.
\end{equation}
\end{enumerate}
with $E_{i}$ (respectively $E_{j}$) the payoff of a player $i$ (respectively $j$), $\beta=1$ and $M=min\left(T,R,S,P\right)$. 

We have also performed simulations with synchronous and asynchronous updating (which are denoted, respectively, as $s=1$ and $s=0$). The results from asynchronous updating are not reported in the main text but are reported in the supporting materials\cite{GoG:arXiv_Jinshan}, because the results are not very different from the results reported here on synchronous updating.

\section{Results}

All of the networks are of the size $N=51 \times 51$ in our simulations. The average degree $\langle k \rangle$ of the two lattices, the von Neumann neighborhood and the Moore neighborhood, are $4$ and $8$, respectively. The WS model is generated from a $51 \times 51$ square lattice with an average degree of $\langle k \rangle=4$ and a rewiring probability of $q=0.2$. In the BA model, we set the parameter $m=2$, which denotes the number of edges of a new node that is added to the already present networks. The relation between the parameter $m$ and the average degree is $\langle k \rangle=2m$ \cite{barabasi1999emergence}; thus, the average degree $\langle k \rangle$ of the BA model is also $4$.

Each player can be either a cooperator or a defector, and their initial strategic status is determined randomly according to a fraction of the cooperators $f_0$. Two values of $f_0$, $0.2$ and $0.8$ (denoted respectively as initial condition $i=0$ and $i=1$), are used in our simulation. Here, we allow these two free variables $S$ and $T$ to take on arbitrary values from $\left[-5, 5\right]$ with an interval of $0.2$. For a given setup with a fixed $S$, $T$, $f_0$, network, updating rule and payoff scheme, after the simulation is stabilized long enough (after $L$ rounds of evolution), we record the final fraction of cooperators from averaging over a period of time $\Delta L$. The same simulation is repeated $N$ times, and the final fraction of cooperators of the game under the given configuration is then calculated from the average of all $N$ trajectories. We then plot the value of this fraction on the square coordinates of $S$ and $T$. We call this plot a phase diagram. Different games under different configurations require different values of $L$ and $N$; we present such details in the supporting materials\cite{GoG:arXiv_Jinshan}.

\begin{figure}[htb] \centering
\includegraphics[width=0.2\textwidth]{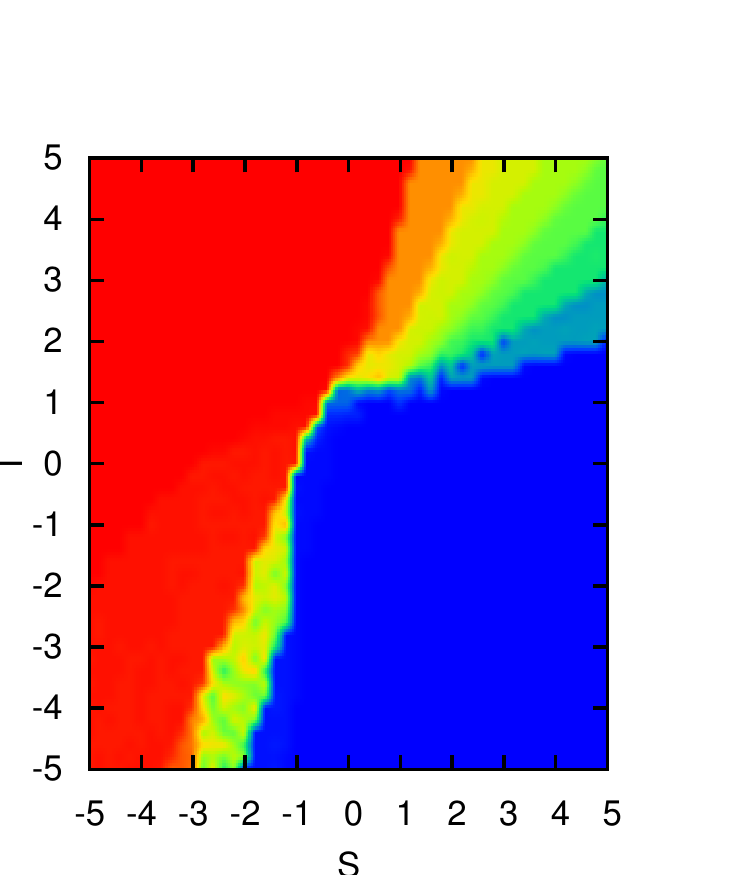}
\includegraphics[width=0.2\textwidth]{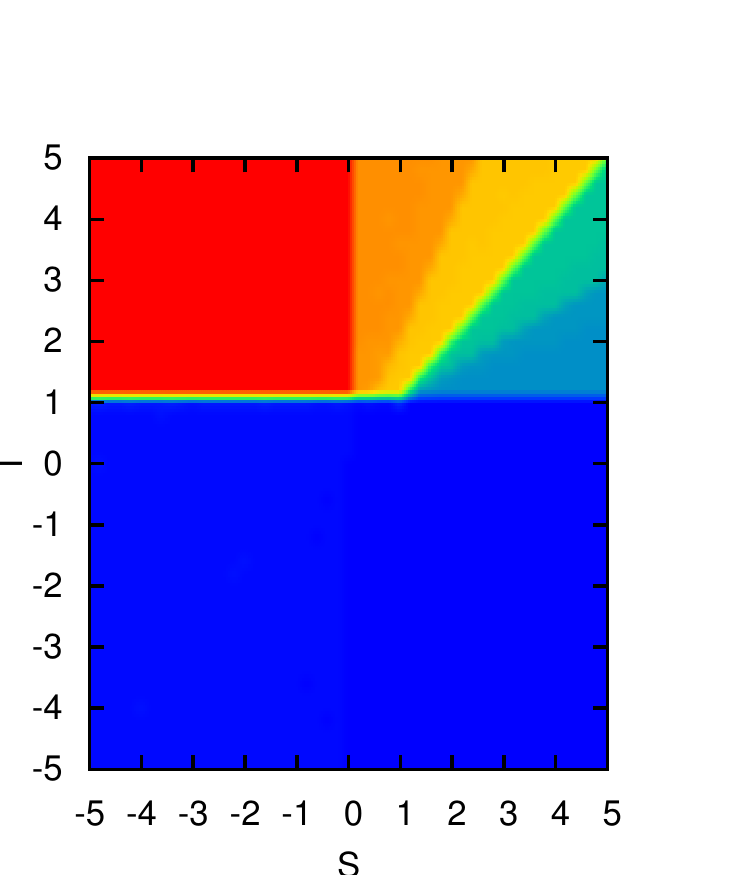}
\includegraphics[width=0.2\textwidth]{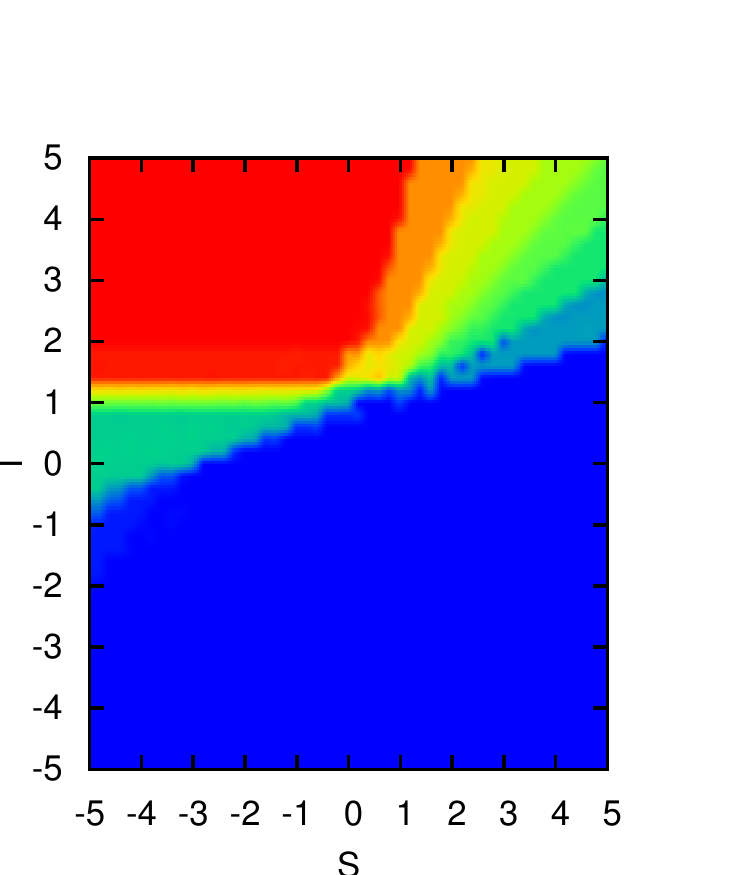}
\includegraphics[width=0.2\textwidth]{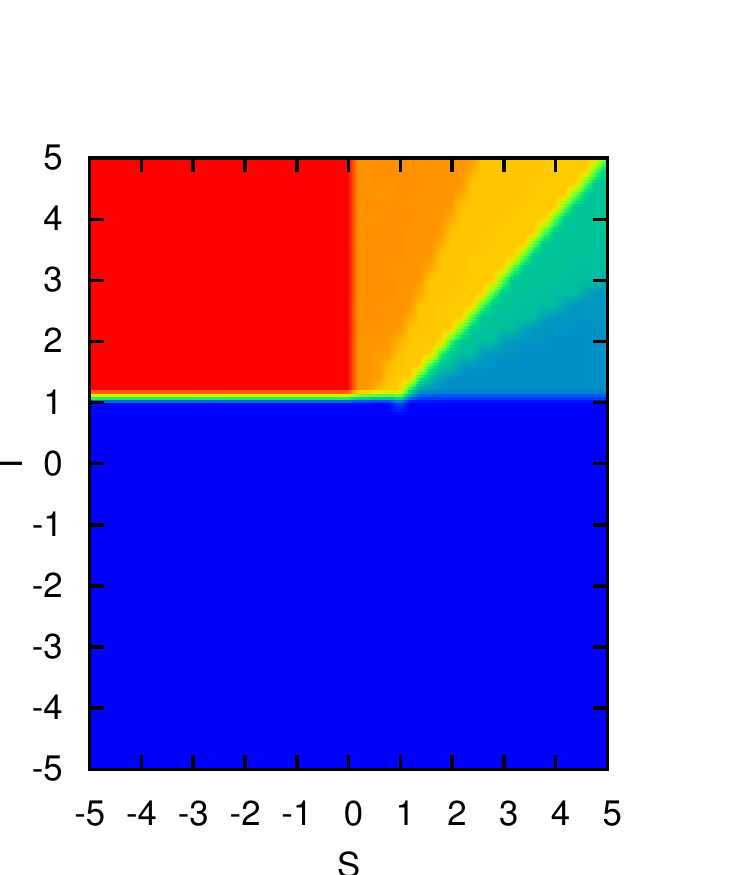}
\caption{\label{fig1} Phase diagrams for synchronous updating, imitating the best on a von-Neumann-neighborhood lattice. (a) Average Payoff, $f_{0}=0.2$, (b) Stochastic Payoff, $f_{0}=0.2$, (c) Average Payoff, $f_{0}=0.8$, and (d) Stochastic Payoff, $f_{0}=0.8$. The colors changing from blue to red correspond to the levels of cooperation from $1$ to $0$.}
\end{figure}

Here, we first show several examples of the phase diagrams, which show certain typical features of our observations. A comparison of all of these phase diagrams will be discussed later, while all of our other phase diagrams are provided in the supporting materials\cite{GoG:arXiv_Jinshan}. In Fig.\ref{fig1}, we plot the phase diagram, which shows $f_{c}$ as a function of $S$ and $T$ with a synchronous updating rule, mimicking the best, on a von-Neumann neighborhood, and using the initial fractions of cooperators $f_{0}=0.2$ and $f_{0}= 0.8$. Fig.\ref{fig1}(a) and Fig.\ref{fig1}(c) have been reported on elsewhere \cite{hauert2002effects}. We have regenerated and confirmed the results from our own simulations. Fig.\ref{fig1}(b) and Fig.\ref{fig1}(d) show the results from using a stochastic payoff scheme.

As can be clearly seen, the equilibrium cooperation levels are notably different between the average payoff scheme and the stochastic payoff scheme. In fact, when comparing between the payoff schemes, the stochastic payoff scheme appears to facilitate more cooperation regardless of the initial cooperative fraction. This observation holds especially in the area defined by $S<0,T<1$. Fig. \ref{fig1} also shows that when a stochastic payoff scheme is used, the level of cooperation is the same for the different initial conditions $f_{0}=0.2$ (Fig. \ref{fig1}(b)) and $f_{0}=0.8$ (Fig.\ref{fig1}(d)), while in the case of an average payoff scheme, different initial conditions lead to qualitatively different levels of cooperation.  

\begin{figure}[htb] \centering
\includegraphics[width=0.2\textwidth]{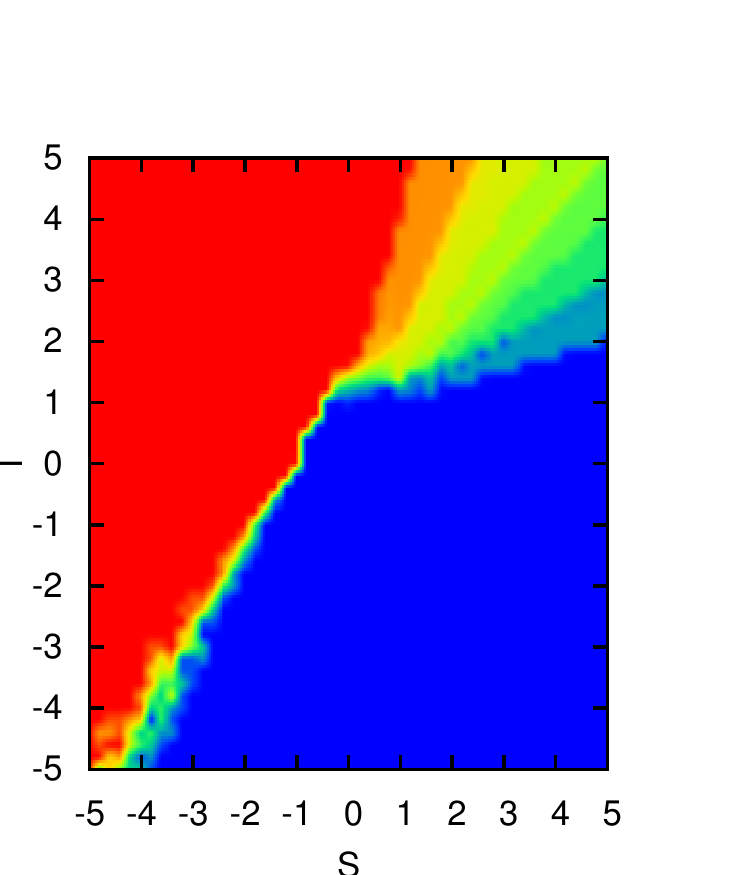}
\includegraphics[width=0.2\textwidth]{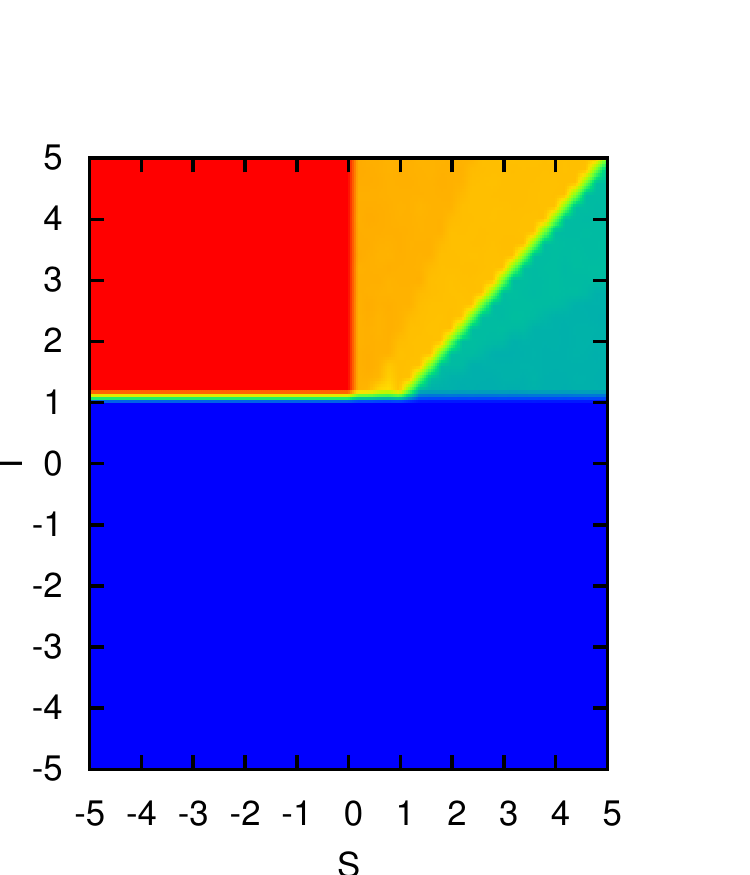}
\includegraphics[width=0.2\textwidth]{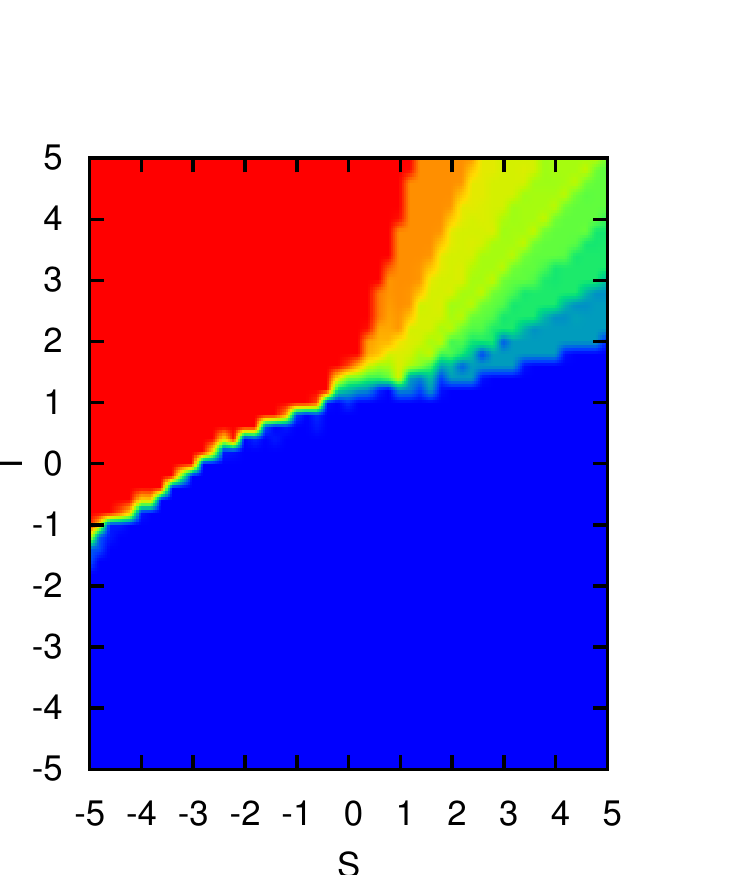}
\includegraphics[width=0.2\textwidth]{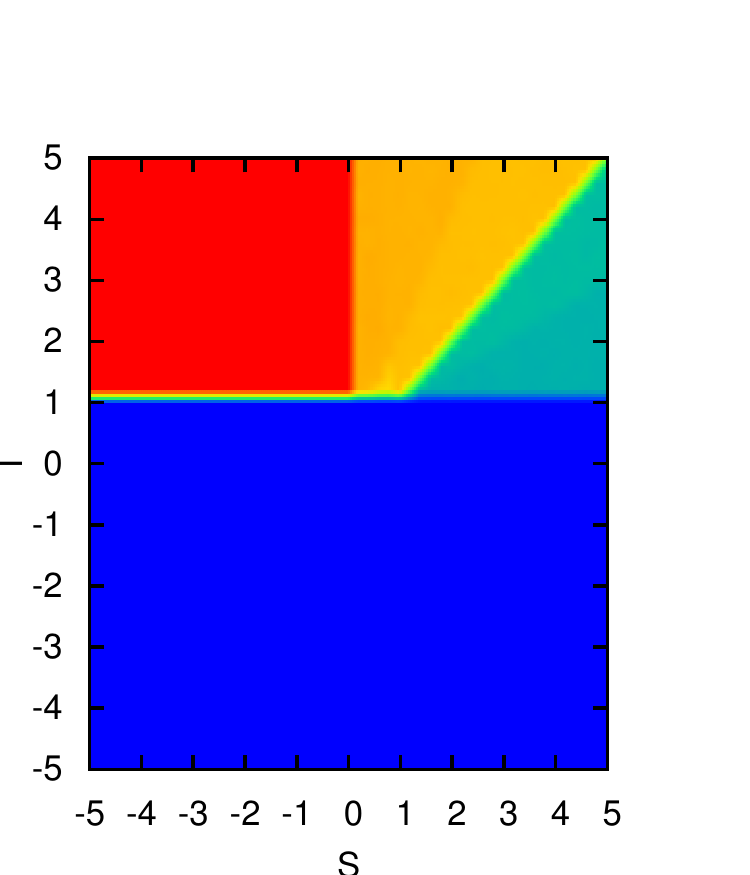}
\caption{\label{fig2} Phase diagrams for synchronous updating, imitating the better linearly with a von-Neumann-neighborhood lattice. All of the other configurations are the same as those in Fig. \ref{fig1}.}
\end{figure}

To provide another example, we also plot here in Fig. \ref{fig2} the corresponding phase diagrams for synchronous updating, imitating the better with a linear probability on a von-Neumann-neighborhood lattice. Qualitatively, Fig. \ref{fig2} demonstrates the same features of our observations that the difference due to different payoff schemes is quite visible and the initial level of cooperation makes almost no difference in the stochastic scheme while it does make a difference in the average scheme, plus level of cooperation is higher when the stochastic scheme other than the average scheme is used.

We have generated phase diagrams for all $80$ different settings, including $2$ payoff schemes (denoted as $p=0, 1$), $5$ updating rules (denoted as $u=0, 1, 2, 3,$ and $4$), $4$ networks (denoted as $n=0, 1, 2,$ and $3$) and $2$ initial conditions (denoted as $i=0$ and $1$). A game with a specific setting is denoted as $g_{puni}=g_{40p+8u+2n+i}$, {\it{i.e.}}, game No. $\left(40p+8u+2n+i\right)$. 

To demonstrate the two major observations that the two payoff schemes lead to qualitatively different levels of cooperation and behavior from a scholastic scheme that is more robust, in the following we compare all of the $80$ games to one another. A comparison of the entire $160$ games, including both synchronous (denoted as $s=1$) and asynchronous (denoted as $s=0$) updating, can be found in the supporting materials\cite{GoG:arXiv_Jinshan}.

\begin{figure}[htb] \centering
\includegraphics[width=0.45\textwidth]{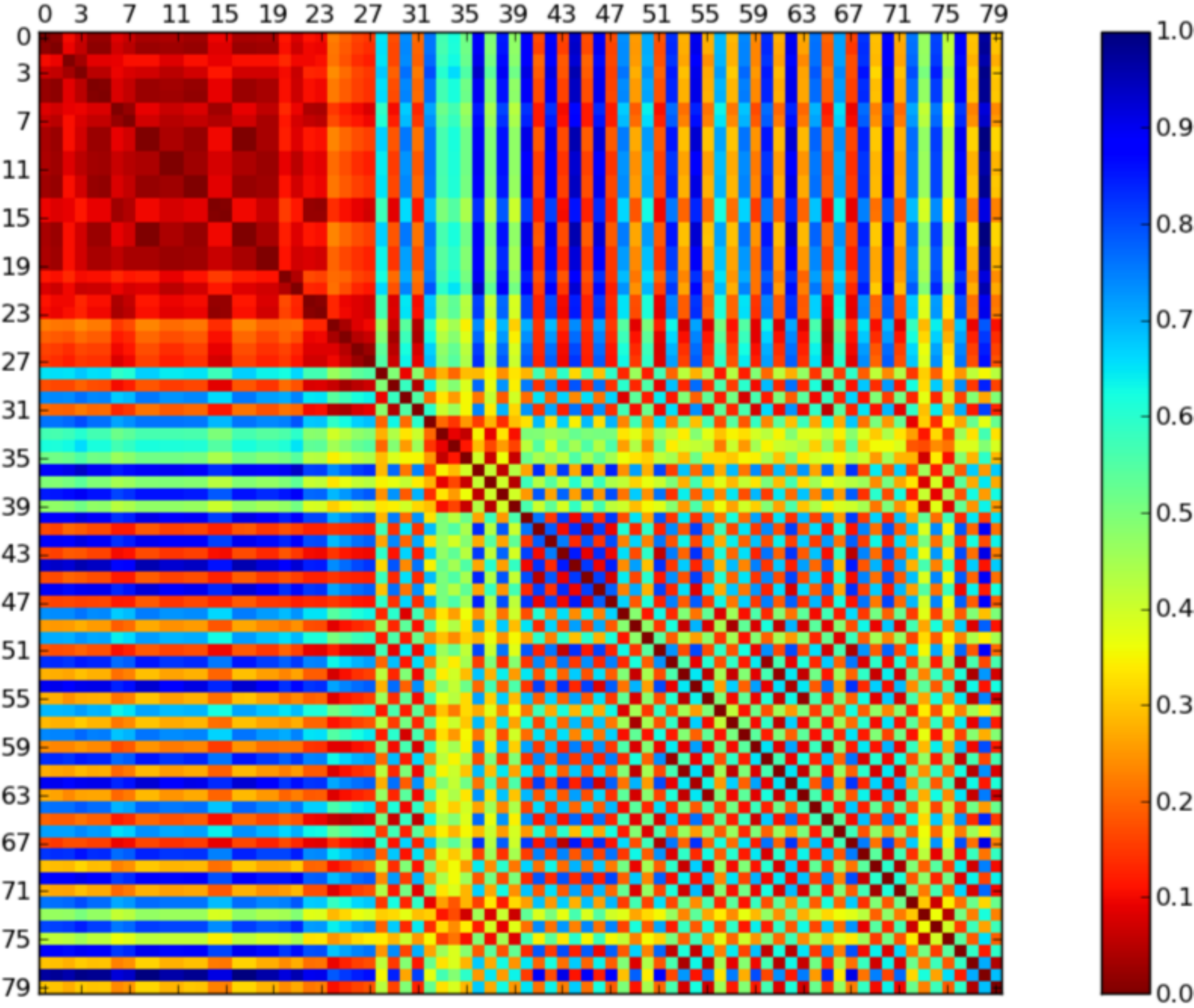}
\caption{\label{fig3} Comparison of all $80$ configurations. The first (latter) $40$ games use the stochastic (average) payoff scheme. Major observations are the following: (1) The diagonal part of the first $40$ games is relatively small; thus, all of the games (except games under the fifth updating rule --- Proportional update with a linear probability) under the stochastic scheme have a similar level of cooperation; (2) The diagonal part of the latter $40$ games is relatively large; thus, the levels of cooperation are not very similar among those games; (3) The difference between the first and the latter $40$ games (off-diagonal part) are clearly larger than those among the first $40$ games. } 
\end{figure}

In Fig. \ref{fig3}, we compare all $80$ configurations. Each point in the figure, $d_{lm}=\frac{\sum_{i,j} \lbar f_{l}\left(S_{i},T_{j}\right) - f_{m}\left(S_{i},T_{j}\right)\rbar}{d_{max}}$ where $f_{l}\left(S_{i},T_{j}\right)$ is the level of cooperation of game $g_{l}$ at parameter value $S_{i}, T_{j}$, corresponds to the difference in the level of cooperation between game $g_{l}$ and game $g_{m}$. Here $d_{max}=max\left\{d_{lm}\right\}$. We can see that there is a visible difference between the the first (stochastic scheme) and the latter (average scheme) $40$ games (the off-diagonal part between the first and the latter $40$ games). It is also evident that the small differences among the games with a stochastic scheme (the diagonal part of the first $40$ games except games under the fifth updating rule --- Proportional update with a linear probability, which seem to be slight different from results under other updating rules. We do not have intuitive understanding of this difference.) demonstrate that they all have similar levels of cooperation. Thus, the results from the games that have stochastic schemes are robust with regard to all of the other variables. At the same time, differences among the latter $40$ games (the diagonal part of the latter $40$ games), which are under the average scheme, are much larger. Thus, the results from the average scheme are not as robust as the stochastic scheme. Additional results on all phase diagrams on every one of the $160$ combinations of the parameters can be found in \cite{GoG:arXiv_Jinshan}.

\section{Conclusions and Discussions}

Here, we have compared the stochastic scheme against the average scheme. We found that although this modification seems minor, our simulation shows qualitatively different levels of cooperation from the two schemes with regard to all of the other conditions of the games, including the underlying networks, rules of updating, initial level of cooperation and synchronous/asynchronous updating. This finding appears to provide a possible explanation for the wide spectrum of predicted behaviors from various theoretical works in the literature: a seemingly non-essential modification of the mechanisms can lead to a large difference. Furthermore, we observed that across initial conditions and the underlying networks, the levels of cooperation are more robust when the stochastic scheme is used than those of the average scheme. This finding suggests that perhaps in studies of games on graphs, rather than the average scheme, the stochastic scheme should be used. 


The stochastic scheme seems to be very similar with the average scheme but it leads to quite different results: Generally speaking the level of cooperation is higher when the stochastic scheme other than the average is used. We believe that this big difference due to minor modification of payoff scheme is related to the fact that there is not a well-defined energy of the whole system and there is not a detailed balance principle for games systems. In physical systems such as Ising model, there are overall energy functions and the principle of detailed balance thus the order of choosing which spins to flip does not matter.

We also believe that the stochastic scheme makes better sense in modeling a real-life game-playing experience. Unless there is a central agency of mediators, the cost, in terms of attention or resources, of a player interacting with all of its neighbors at each round of a game, increases with the number of neighboring players. Therefore, it is reasonable to assume that, in each round, the player will reach to only one or a few reachable players, but not to all of its neighbors.

\acknowledgments
The authors wish to thank Christoph Hauert for sharing his data (Fig. \ref{fig1}(a)) and for insightful discussion on various issues about our results. This work was supported by the Fundamental Research Funds for the Central Universities of China. The authors also like to thank the anonymous referees for their very insightful comments, which have made several of the observations in this work clearer.

\section{Appendix}
Here we provide further details on our simulation and also include all the generated figures for the whole $160$ cases of values of the parameters.

\subsection{Further details on the implementation of our simulation}
Besides the underlying $2\times 2$ payoff matrices, which are determined by $S$ and $T$ because $R=1, P=0$, a game that is fully described by the configuration $psuni$. $puni$ has been discussed in the main text. Here, the additional $s$ represents synchronized ($s=1$) or asynchronous ($s=0$) updating. Then, the whole $160$ games are ordered as $g_{psuni}=g_{80p+40(s-1)+8u+2n+i}$.
  
\begin{figure}[htb] \centering
\includegraphics[width=0.4\textwidth]{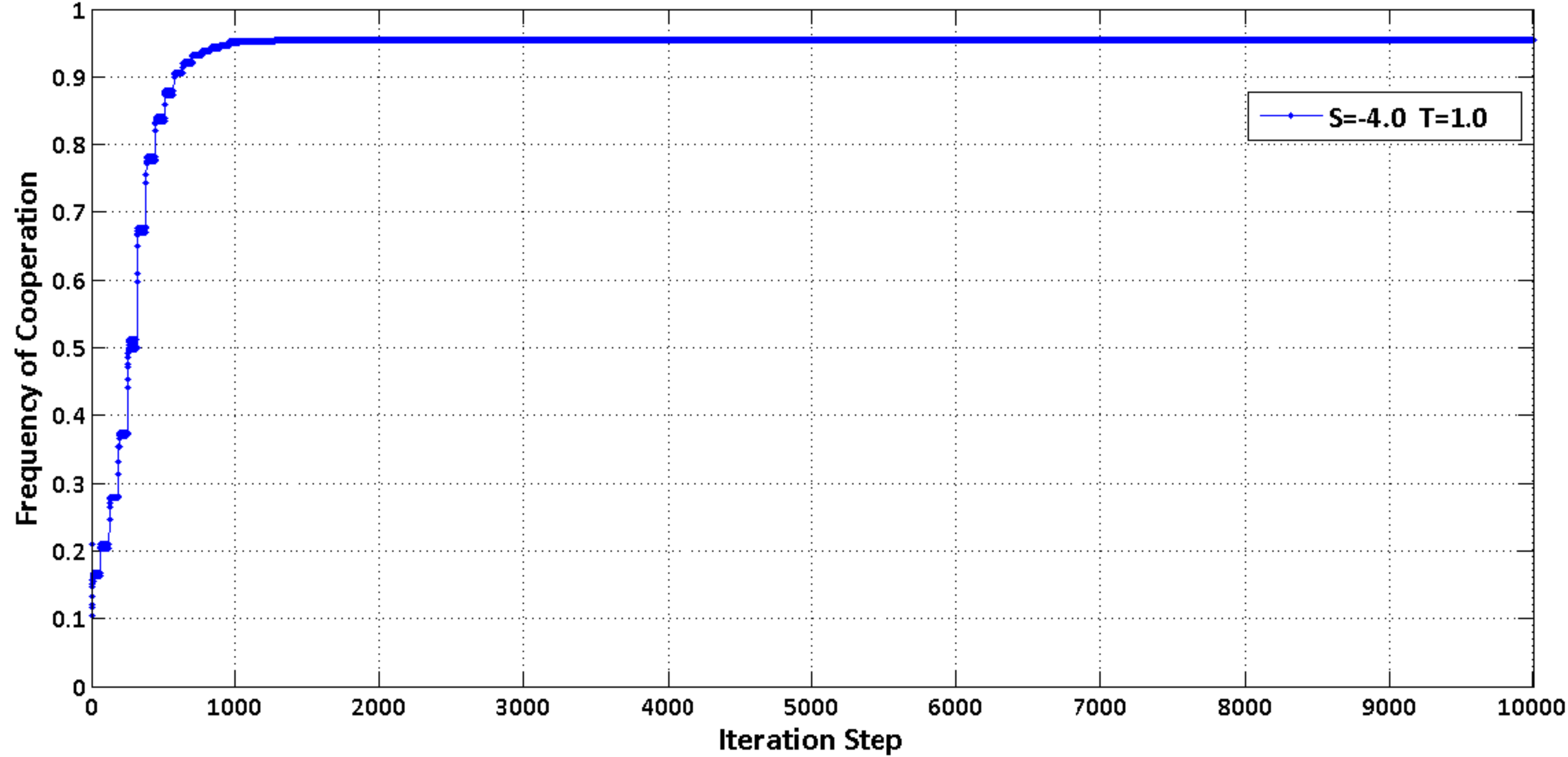}
\caption{\label{figs1} For a game with a given setup of $psuni=(0,0,0,2,0)$, which means that it has a WS network, asynchronous updating, a stochastic payoff scheme, imitating the best, $f_0=0.2$, with $S=-4.0$ and $T=1.0$, over one realization of the WS network, we run the simulation long enough to obtain $f_{c}$, the equilibrium level of cooperation.}
\end{figure}

Given the game's configuration, we are interested in the equilibrium frequency of cooperation, $f_{c}$, as a function of the parameters $S$ and $T$. In our simulation, $f_{c}$ is obtained by averaging over the last $300$ iterations of the entire $10000$ iterations, which we will show in the following is long enough to reach equilibrium. For each setup, including the specified values of $psuni$ and $S, T$, in the case of the WS networks and the BA networks, we run the simulation on $20$ different network realizations. Given all of the setup and network structures, we perform $30$ runs for each realization using different initial conditions, however, with the same initial level of cooperation. We have confirmed that those chosen parameters are sufficiently large that averaging over larger periods or more realizations does not lead to noticeable changes in the value of $f_{c}$. For example, in Fig. \ref{figs1}, we plot the time evolution of one game with a specified setup of $psuni=(0,0,0,2,0)$, $S=-4.0$ and $T=1.0$ on a single realization of the WS network. We can see that, long before $10000$ iterations, it already reaches equilibrium. Other choices of parameters for our simulation have been similarly tested. 

We have also tested effects of finite size. Here we set the total number of players on every graph to be $N=2601=51\times 51$. We have compared the level of cooperation to larger networks such as $N=80\times 80$ and $N=100\times 100$ and we found no visiable difference. The parameter $N=2601=51\times 51$ is used in \cite{hauert2002effects}. Other simulations\cite{chen2008promotion, du2009evolutionary, chen2010evolutionary} usually user smaller number of players.

Next, we report all of the results on the equilibrium levels of cooperation of each of the $160$ configurations. To make all of the figures have the same dimension and, thus, be easier for human eyes, we also include those two that have already been covered in the main text. 

\subsection{Phase diagrams for all $80$ games with synchronized updating}
Let us start first from the synchronized updated games.
 
\begin{figure}[htb] 
\includegraphics[width=0.11\textwidth]{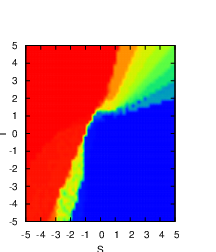}
\includegraphics[width=0.11\textwidth]{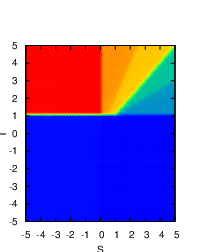}
\includegraphics[width=0.11\textwidth]{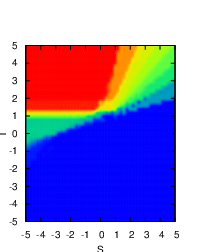}
\includegraphics[width=0.11\textwidth]{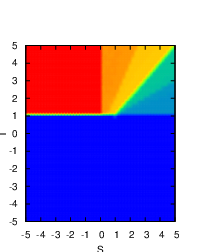}

\includegraphics[width=0.11\textwidth]{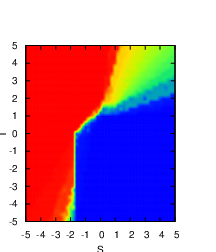}
\includegraphics[width=0.11\textwidth]{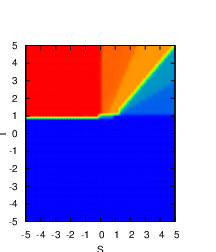}
\includegraphics[width=0.11\textwidth]{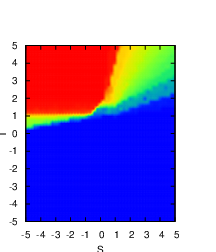}
\includegraphics[width=0.11\textwidth]{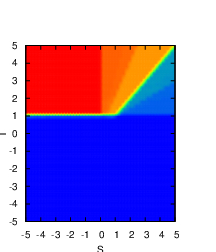}

\includegraphics[width=0.11\textwidth]{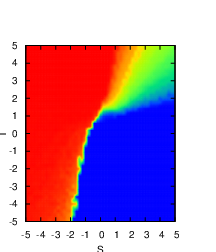}
\includegraphics[width=0.11\textwidth]{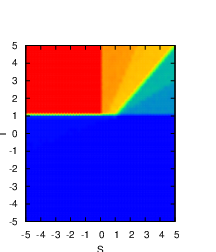}
\includegraphics[width=0.11\textwidth]{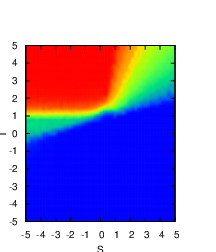}
\includegraphics[width=0.11\textwidth]{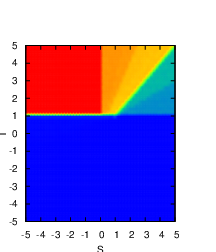}

\includegraphics[width=0.11\textwidth]{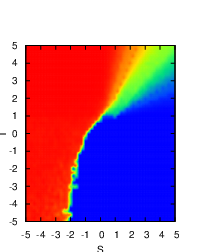}
\includegraphics[width=0.11\textwidth]{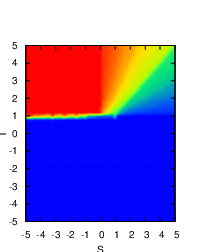}
\includegraphics[width=0.11\textwidth]{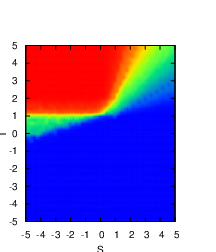}
\includegraphics[width=0.11\textwidth]{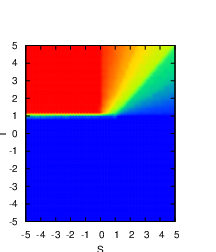}
\caption{\label{figs2} Phase diagrams for games with synchronized updating and imitating the best. Different rows from top to bottom correspond to different networks: von-Neumann-neighborhood lattice, Moore-neighborhood lattice, Watts-Strogatz small-world and BA scale-free networks. Different columns from left to right refer to the following: (1) Average Payoff, $f_{0}=0.2$, (2) Stochastic Payoff, $f_{0}=0.2$, (3) Average Payoff, $f_{0}=0.8$, and (4) Stochastic Payoff, $f_{0}=0.8$. The colors changing from blue to red correspond to the levels of cooperation from $1$ to $0$.}
\end{figure}

From Fig. \ref{figs2}, we find that the first results from the stochastic scheme are obviously different from those from the average scheme, and the second network structures and initial conditions do not have much influence on the levels of cooperation for games under the stochastic scheme, while they do have visible impact on games under the average scheme.

\begin{figure}[htb] 
\includegraphics[width=0.11\textwidth]{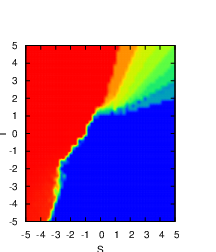}
\includegraphics[width=0.11\textwidth]{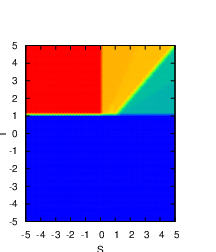}
\includegraphics[width=0.11\textwidth]{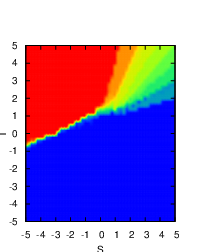}
\includegraphics[width=0.11\textwidth]{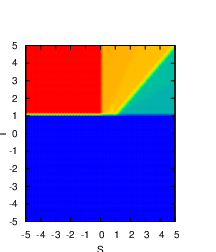}

\includegraphics[width=0.11\textwidth]{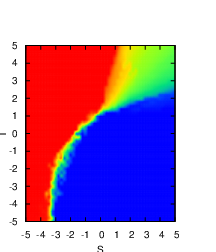}
\includegraphics[width=0.11\textwidth]{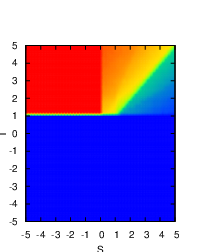}
\includegraphics[width=0.11\textwidth]{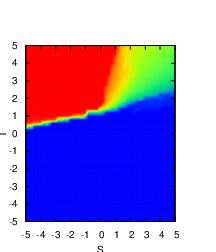}
\includegraphics[width=0.11\textwidth]{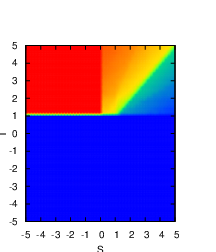}

\includegraphics[width=0.11\textwidth]{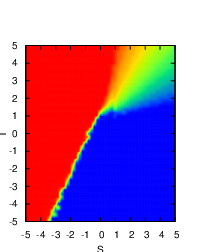}
\includegraphics[width=0.11\textwidth]{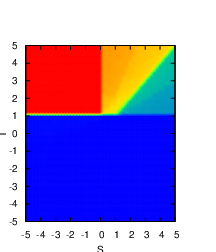}
\includegraphics[width=0.11\textwidth]{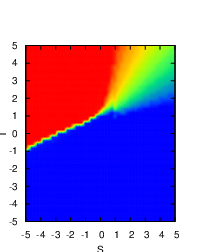}
\includegraphics[width=0.11\textwidth]{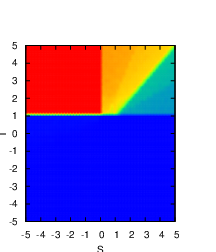}

\includegraphics[width=0.11\textwidth]{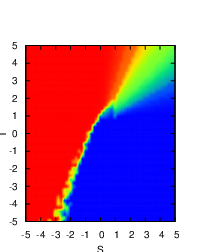}
\includegraphics[width=0.11\textwidth]{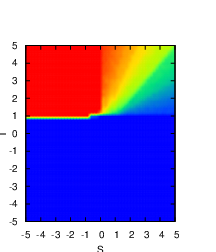}
\includegraphics[width=0.11\textwidth]{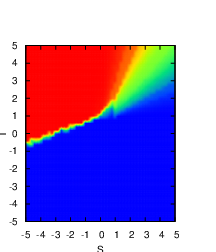}
\includegraphics[width=0.11\textwidth]{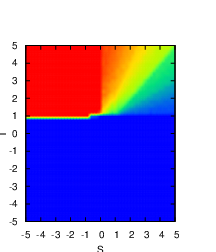}
\caption{\label{figs3} Phase diagrams for synchronized updating and imitating the better exponentially. The same orders and parameters are used as in Fig.\ref{figs2}.}
\end{figure}

\begin{figure}[htb]
\includegraphics[width=0.11\textwidth]{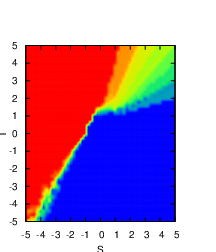}
\includegraphics[width=0.11\textwidth]{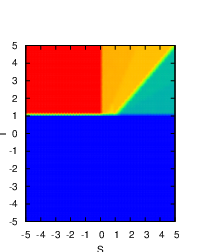}
\includegraphics[width=0.11\textwidth]{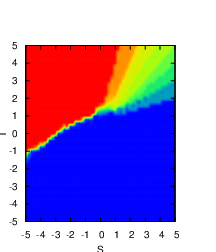}
\includegraphics[width=0.11\textwidth]{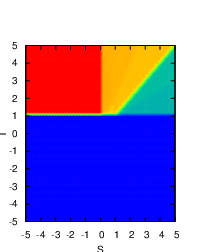}

\includegraphics[width=0.11\textwidth]{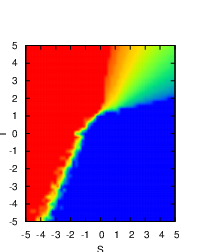}
\includegraphics[width=0.11\textwidth]{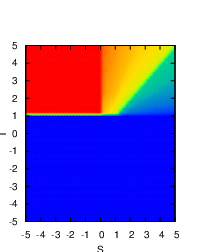}
\includegraphics[width=0.11\textwidth]{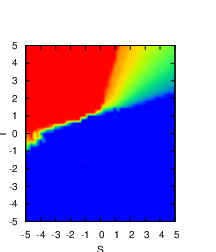}
\includegraphics[width=0.11\textwidth]{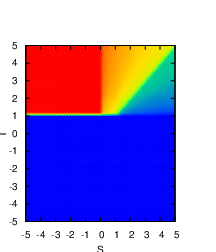}

\includegraphics[width=0.11\textwidth]{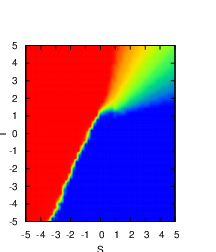}
\includegraphics[width=0.11\textwidth]{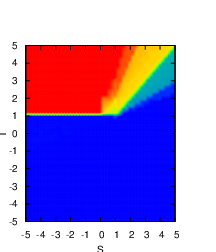}
\includegraphics[width=0.11\textwidth]{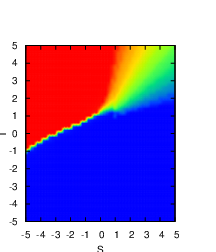}
\includegraphics[width=0.11\textwidth]{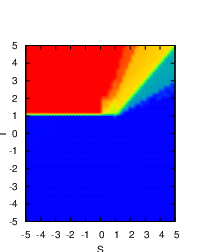}

\includegraphics[width=0.11\textwidth]{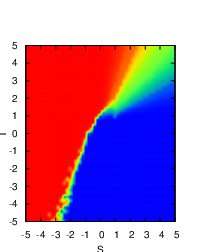}
\includegraphics[width=0.11\textwidth]{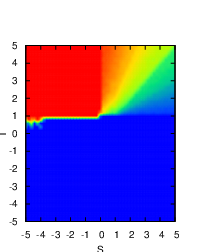}
\includegraphics[width=0.11\textwidth]{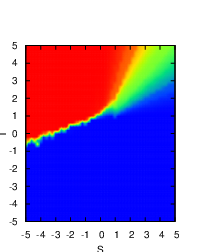}
\includegraphics[width=0.11\textwidth]{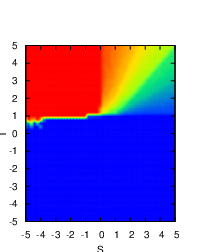}
\caption{\label{figs4} Phase diagrams for synchronized updating and imitating the better with a linear probability. The same orders and parameters are used as in Fig.\ref{figs2}.}
\end{figure}

From Figs. \ref{figs3} and \ref{figs4}, again we find that the results from the stochastic scheme are obviously different from those from the average scheme. For the games under the stochastic scheme, both the network structure and the initial conditions do not have much influence on the levels of cooperation. For games under the average scheme, different initial conditions lead to visible differences, while the results are not very sensitive to different networks.

\begin{figure}[htb] 
\includegraphics[width=0.11\textwidth]{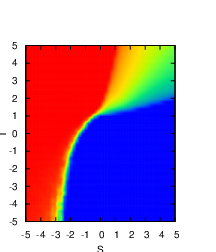}
\includegraphics[width=0.11\textwidth]{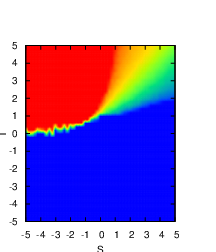}
\includegraphics[width=0.11\textwidth]{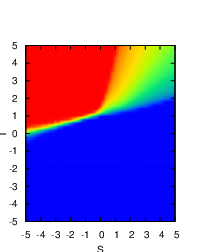}
\includegraphics[width=0.11\textwidth]{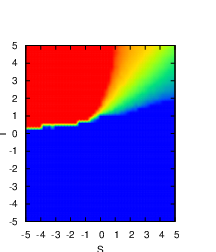}

\includegraphics[width=0.11\textwidth]{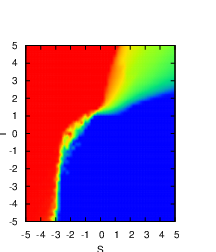}
\includegraphics[width=0.11\textwidth]{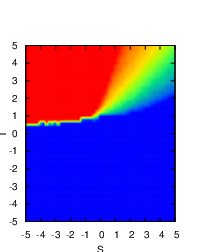}
\includegraphics[width=0.11\textwidth]{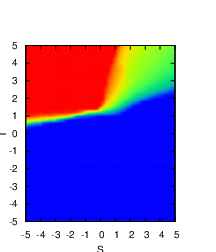}
\includegraphics[width=0.11\textwidth]{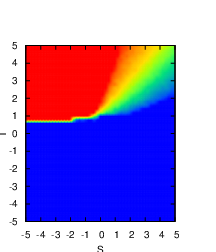}

\includegraphics[width=0.11\textwidth]{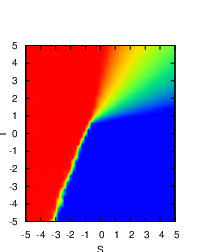}
\includegraphics[width=0.11\textwidth]{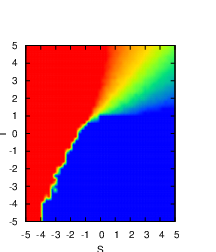}
\includegraphics[width=0.11\textwidth]{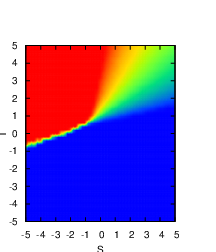}
\includegraphics[width=0.11\textwidth]{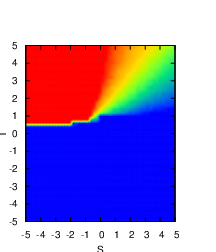}

\includegraphics[width=0.11\textwidth]{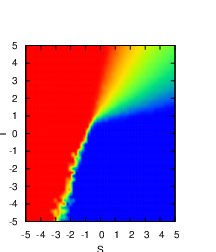}
\includegraphics[width=0.11\textwidth]{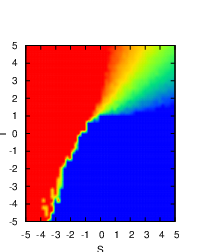}
\includegraphics[width=0.11\textwidth]{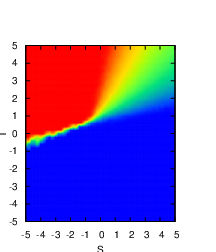}
\includegraphics[width=0.11\textwidth]{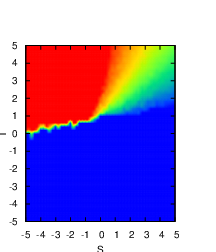}
\caption{\label{figs5} Phase-plane diagrams for synchronized updating and proportional update rules with exponential probabilities. The same orders and parameters are used as in Fig.\ref{figs2}.}
\end{figure}

In Fig. \ref{figs5}, again we find that the results from the stochastic scheme are obviously different from those from the average scheme. On the robustness of the results from the stochastic scheme, Fig. \ref{figs5} is slightly different from the previous $3$ figures in that on the WS and BA networks, the initial conditions ($f_{0}=0.2$ and $f_{0}=0.8$) result in noticeably different levels of cooperation. Except for these two cases, all of the other games have similar phase diagrams.

\begin{figure}[htb] 
\includegraphics[width=0.11\textwidth]{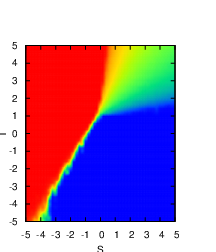}
\includegraphics[width=0.11\textwidth]{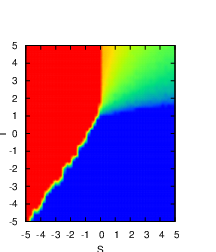}
\includegraphics[width=0.11\textwidth]{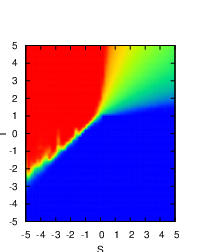}
\includegraphics[width=0.11\textwidth]{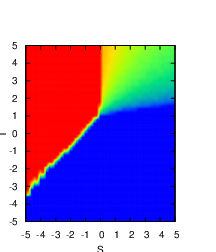}

\includegraphics[width=0.11\textwidth]{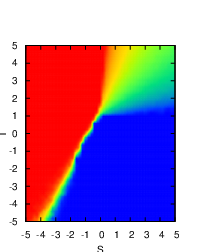}
\includegraphics[width=0.11\textwidth]{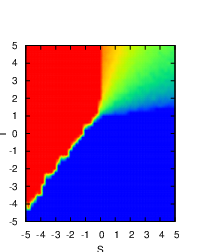}
\includegraphics[width=0.11\textwidth]{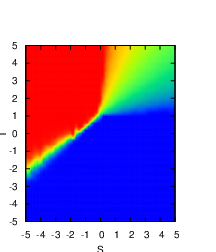}
\includegraphics[width=0.11\textwidth]{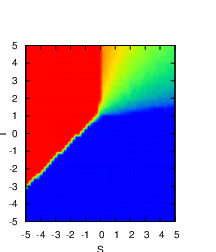}

\includegraphics[width=0.11\textwidth]{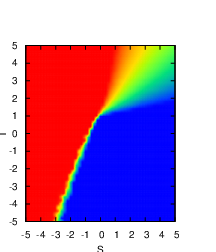}
\includegraphics[width=0.11\textwidth]{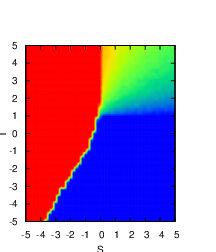}
\includegraphics[width=0.11\textwidth]{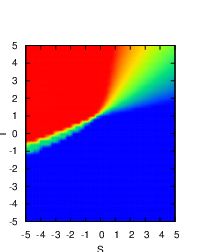}
\includegraphics[width=0.11\textwidth]{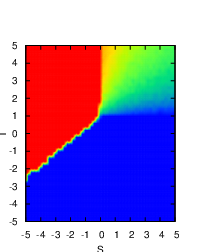}

\includegraphics[width=0.11\textwidth]{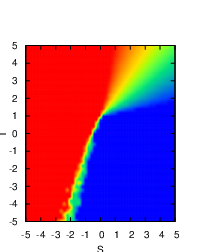}
\includegraphics[width=0.11\textwidth]{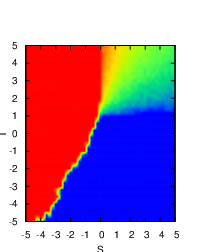}
\includegraphics[width=0.11\textwidth]{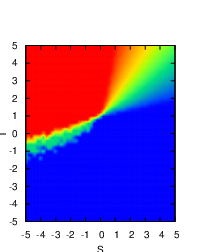}
\includegraphics[width=0.11\textwidth]{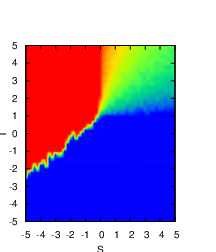}

\caption{\label{figs6} Phase diagrams for the synchronized update and proportional update rule with a linear probability. The same orders and parameters are used as in Fig.\ref{figs2}.}
\end{figure}

Fig. \ref{figs6} is special in that results from the stochastic scheme are not very different from those from the average scheme, except in the case of BA networks. All of the phase diagrams with this setup are more or less similar, and thus, are still robust, while the phase diagrams are visibly different from all of the other cases in the previous $4$ figures. We are not yet clear why different results emerge from these two very similar rules (the proportional update rule with exponential probability in this figure and the proportional update rule with a linear probability in Fig. \ref{figs5}).

\subsection{Phase diagrams for all $80$ games with asynchronous updating}
Next, we provide phase diagrams for all of the asynchronously updated games.

\begin{figure}[htb] 
\includegraphics[width=0.11\textwidth]{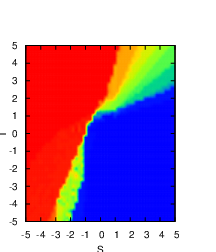}
\includegraphics[width=0.11\textwidth]{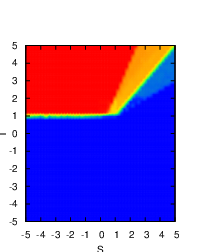}
\includegraphics[width=0.11\textwidth]{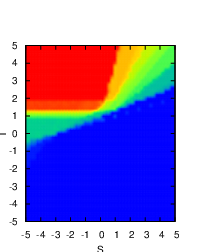}
\includegraphics[width=0.11\textwidth]{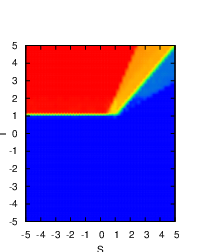}

\includegraphics[width=0.11\textwidth]{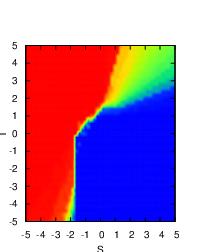}
\includegraphics[width=0.11\textwidth]{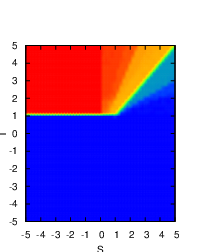}
\includegraphics[width=0.11\textwidth]{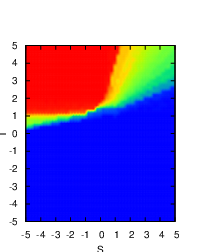}
\includegraphics[width=0.11\textwidth]{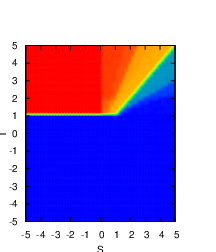}

\includegraphics[width=0.11\textwidth]{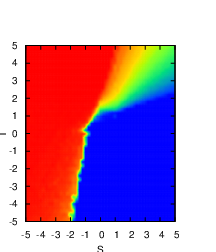}
\includegraphics[width=0.11\textwidth]{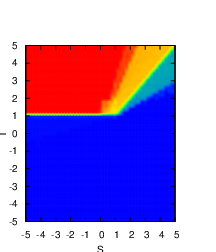}
\includegraphics[width=0.11\textwidth]{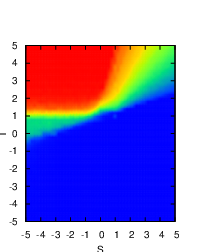}
\includegraphics[width=0.11\textwidth]{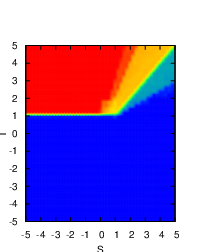}

\includegraphics[width=0.11\textwidth]{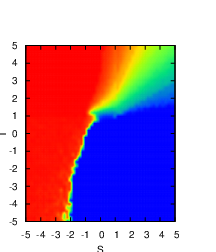}
\includegraphics[width=0.11\textwidth]{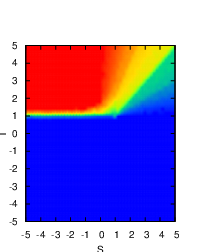}
\includegraphics[width=0.11\textwidth]{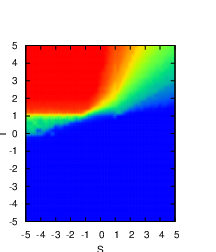}
\includegraphics[width=0.11\textwidth]{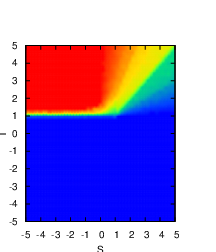}
\caption{\label{figs7} Phase diagrams for asynchronous updating and imitating the best. The same orders and parameters are used as in Fig.\ref{figs2}.}
\end{figure}

In Fig. \ref{figs7}, again we find that results from the stochastic scheme are obviously different from those from the average scheme. Additionally, the results from the stochastic scheme are more robust then those from the average scheme. Furthermore, if we compare this figure with Fig. \ref{figs2}, then we can see that levels of cooperation from those asynchronous games are similar to those from synchronized games. This similarity between synchronized games and asynchronous games is valid for all of the cases of our simulation. Next, we will simply list all of the figures on asynchronous games, and the observations from the figures will be the same as the figures on the corresponding synchronized games.

\begin{figure}[htb] 
\includegraphics[width=0.11\textwidth]{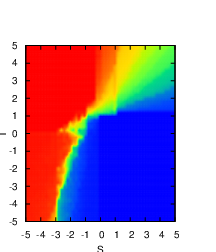}
\includegraphics[width=0.11\textwidth]{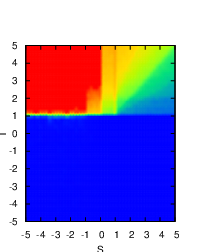}
\includegraphics[width=0.11\textwidth]{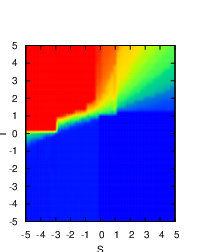}
\includegraphics[width=0.11\textwidth]{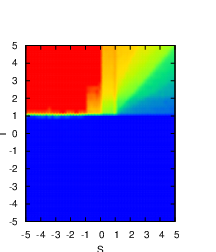}

\includegraphics[width=0.11\textwidth]{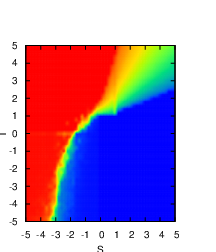}
\includegraphics[width=0.11\textwidth]{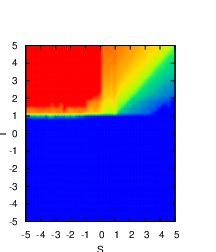}
\includegraphics[width=0.11\textwidth]{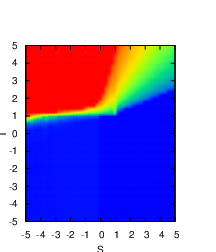}
\includegraphics[width=0.11\textwidth]{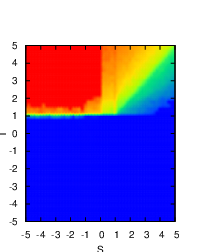}

\includegraphics[width=0.11\textwidth]{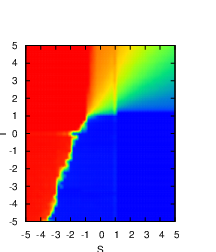}
\includegraphics[width=0.11\textwidth]{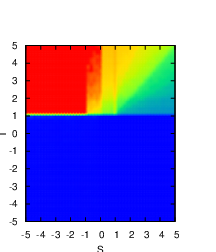}
\includegraphics[width=0.11\textwidth]{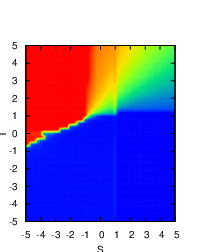}
\includegraphics[width=0.11\textwidth]{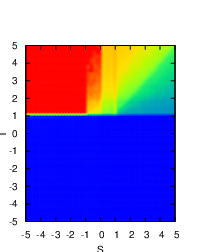}

\includegraphics[width=0.11\textwidth]{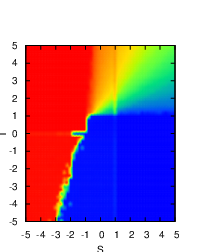}
\includegraphics[width=0.11\textwidth]{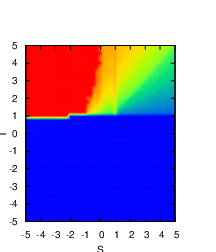}
\includegraphics[width=0.11\textwidth]{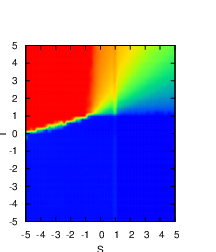}
\includegraphics[width=0.11\textwidth]{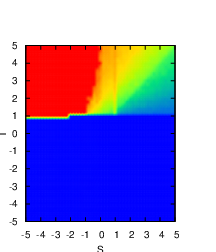}
\caption{\label{figs8} Phase diagrams for asynchronous updating and imitating the better with an exponential probability. The same orders and parameters are used as in Fig.\ref{figs2}.}
\end{figure}

\begin{figure}[htb] 
\includegraphics[width=0.11\textwidth]{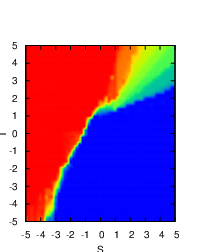}
\includegraphics[width=0.11\textwidth]{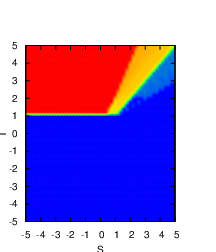}
\includegraphics[width=0.11\textwidth]{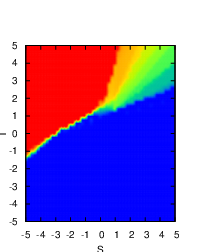}
\includegraphics[width=0.11\textwidth]{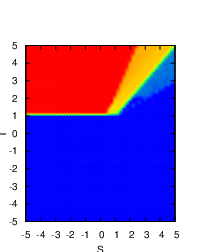}

\includegraphics[width=0.11\textwidth]{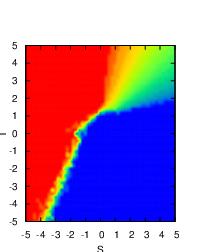}
\includegraphics[width=0.11\textwidth]{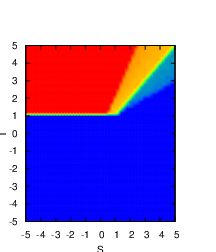}
\includegraphics[width=0.11\textwidth]{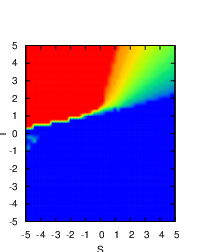}
\includegraphics[width=0.11\textwidth]{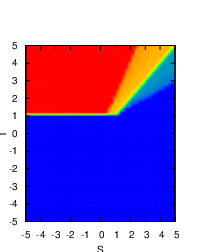}

\includegraphics[width=0.11\textwidth]{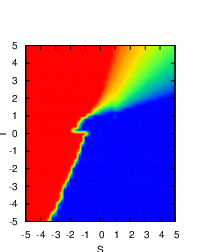}
\includegraphics[width=0.11\textwidth]{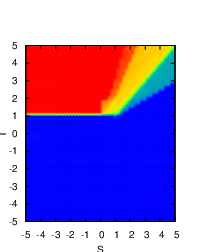}
\includegraphics[width=0.11\textwidth]{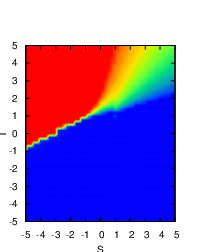}
\includegraphics[width=0.11\textwidth]{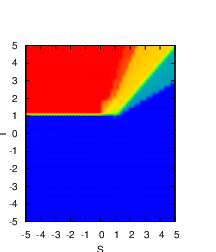}

\includegraphics[width=0.11\textwidth]{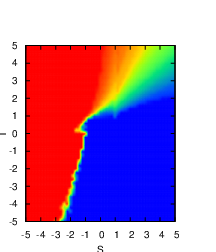}
\includegraphics[width=0.11\textwidth]{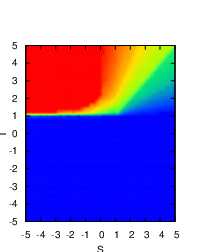}
\includegraphics[width=0.11\textwidth]{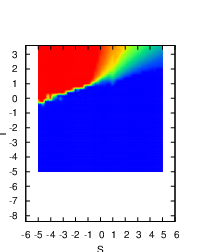}
\includegraphics[width=0.11\textwidth]{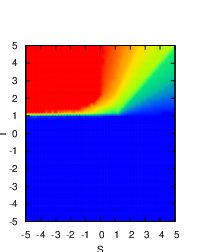}
\caption{\label{figs9} Phase diagrams for asynchronous updating and imitating the better with a linear probability. The same orders and parameters are used as in Fig.\ref{figs2}.}
\end{figure}

\begin{figure}[htb] 
\includegraphics[width=0.11\textwidth]{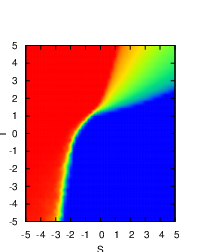}
\includegraphics[width=0.11\textwidth]{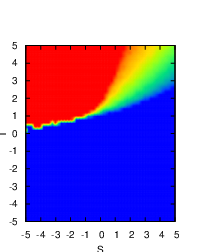}
\includegraphics[width=0.11\textwidth]{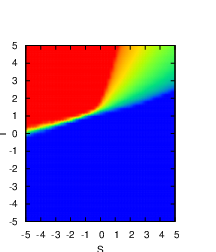}
\includegraphics[width=0.11\textwidth]{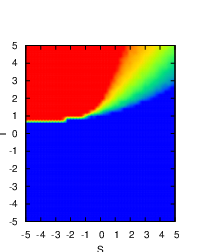}

\includegraphics[width=0.11\textwidth]{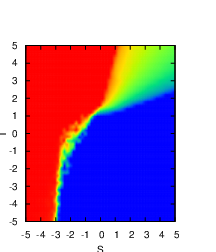}
\includegraphics[width=0.11\textwidth]{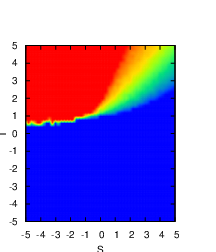}
\includegraphics[width=0.11\textwidth]{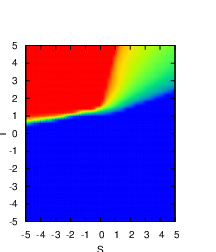}
\includegraphics[width=0.11\textwidth]{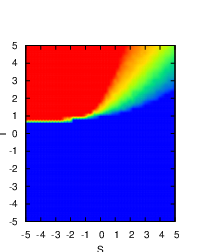}

\includegraphics[width=0.11\textwidth]{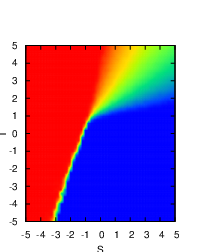}
\includegraphics[width=0.11\textwidth]{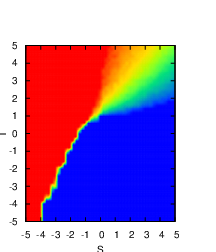}
\includegraphics[width=0.11\textwidth]{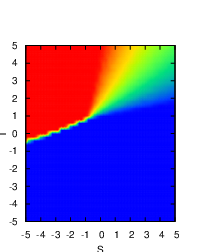}
\includegraphics[width=0.11\textwidth]{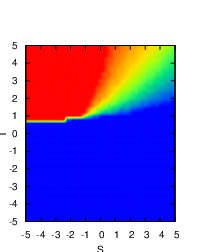}

\includegraphics[width=0.11\textwidth]{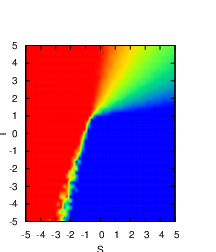}
\includegraphics[width=0.11\textwidth]{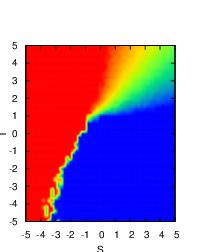}
\includegraphics[width=0.11\textwidth]{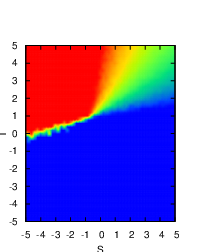}
\includegraphics[width=0.11\textwidth]{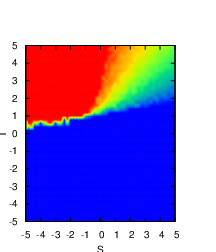}
\caption{\label{figs10} Phase-plane diagrams for asynchronous updating and the proportional updating rule with an exponential probability. The same orders and parameters are used as in Fig.\ref{figs2}.}
\end{figure}

\begin{figure}[htb] 
\includegraphics[width=0.11\textwidth]{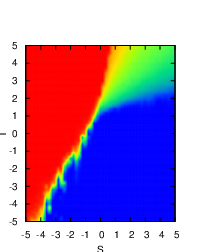}
\includegraphics[width=0.11\textwidth]{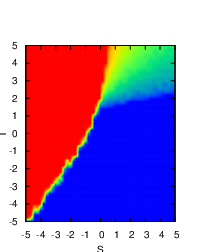}
\includegraphics[width=0.11\textwidth]{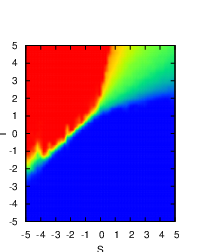}
\includegraphics[width=0.11\textwidth]{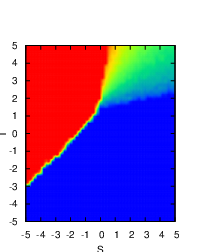}

\includegraphics[width=0.11\textwidth]{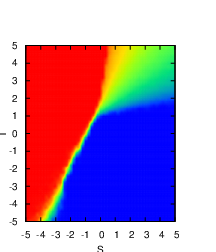}
\includegraphics[width=0.11\textwidth]{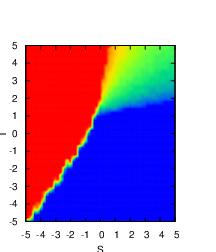}
\includegraphics[width=0.11\textwidth]{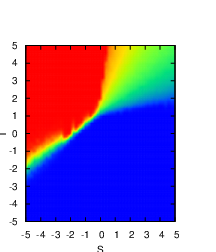}
\includegraphics[width=0.11\textwidth]{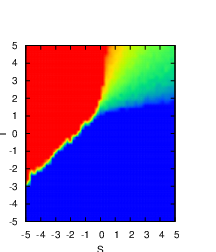}

\includegraphics[width=0.11\textwidth]{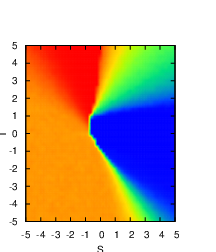}
\includegraphics[width=0.11\textwidth]{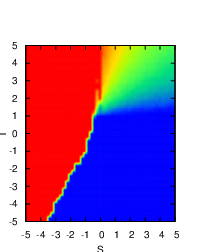}
\includegraphics[width=0.11\textwidth]{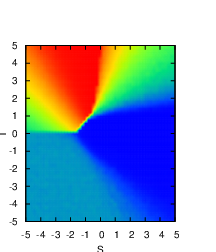}
\includegraphics[width=0.11\textwidth]{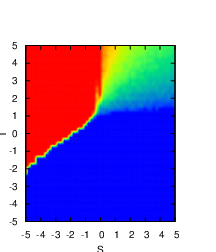}

\includegraphics[width=0.11\textwidth]{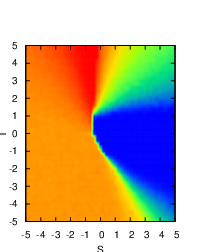}
\includegraphics[width=0.11\textwidth]{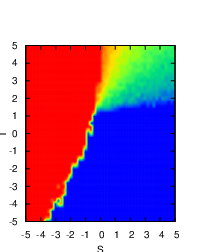}
\includegraphics[width=0.11\textwidth]{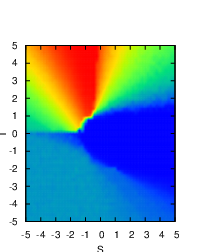}
\includegraphics[width=0.11\textwidth]{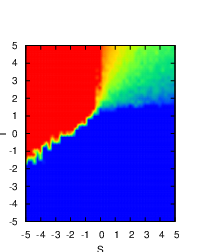}
\caption{\label{figs11} Phase diagrams for the asynchronous update and proportional update rule with a linear probability. The same orders and parameters are used as in Fig.\ref{figs2}.}
\end{figure}

\subsection{Comparison of the $160$ games}
\begin{figure}[htb] \centering
\includegraphics[width=0.45\textwidth]{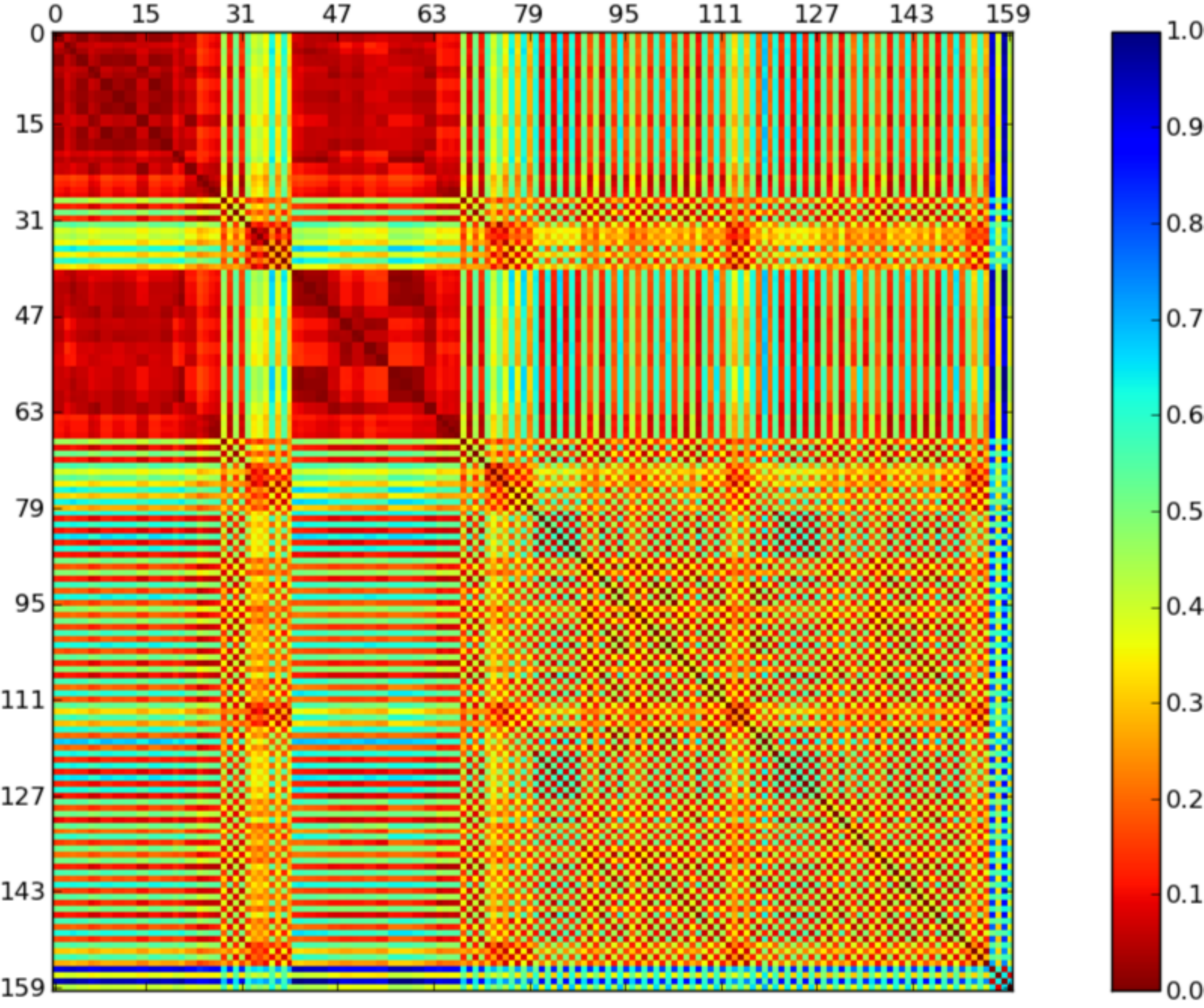}
\caption{\label{figs12} Comparison of all $160$ configurations. The first (latter) $80$ games use the stochastic (average) payoff scheme. Major observations are the following: (1) The diagonal part of the first $60$ games is relatively small; thus, all of the games under the stochastic scheme have similar levels of cooperation; (2) The diagonal part of the latter $80$ games are relatively large; thus, the levels of cooperation are not very similar among those games; (3) The difference between the first and the latter $80$ games are clearly larger than those among the first $80$ games. Thus, the stochastic scheme leads to qualitatively different behavior compared with the results from the average scheme. }
\end{figure}

Similar to Fig. 3 in the main text, in Fig. \ref{figs12} we provide a comparison among all $160$ games. The order of the games has been defined as $l=g_{80p+40(s-1)+8u+2n+i}$ for a game that is specified by $psuni$, and point $d_{lm}$ corresponds to the difference in the level of cooperation between game $g_{l}$ and game $g_{m}$. We can see that there is a visible difference between the first (stochastic scheme) and the latter (average) $80$ games (the off diagonal part between the first and the latter $80$ games). It is also evident that, except for the games with proportional update with a linear probability, the difference among the games with the stochastic scheme (the diagonal part of the first $80$ games) demonstrates similar levels of cooperation. Thus, the results from the games that have stochastic schemes are robust with regard to all of the other variables. At the same time, the differences among the latter $80$ games (the diagonal part of the latter $80$ games), which are under the average scheme, are much larger. Thus, the results for the average scheme are not as robust as for the stochastic scheme. 

\begin{thebibliography}{10}
\expandafter\ifx\csname url\endcsname\relax\def\url#1{\texttt{#1}}\fi

\bibitem{sugden1986economics}
\Name{Sugden R.} \Book{The Economics of Rights, Cooperation and Welfare} (Basil
  Blackwell Oxford) 1986.

\bibitem{sigmund1993games}
\Name{Sigmund K.} \Book{Games of Life: Explorations in Ecology, Evolution and
  Behaviour} (Oxford University Press, Inc.) 1993.

\bibitem{axelrod1981evolution}
\Name{Axelrod R. \and Hamilton W.~D.} \REVIEW{Science}{211}{1981}{1390}.

\bibitem{smith1993evolution}
\Name{Smith J.~M.} \Book{Evolution and the Theory of Games} (Springer) 1993.

\bibitem{weibull1997evolutionary}
\Name{Weibull J.~W.} \Book{Evolutionary Game Theory} (MIT press) 1997.

\bibitem{herbert2000game}
\Name{Gintis H.} \Book{Game Theory Evolving: A Problem-centered Introduction to
  Modeling Strategic Interaction} (Princeton University Press) 2000.

\bibitem{nowak2006evolutionary}
\Name{Nowak M.~A.} \Book{Evolutionary Dynamics: Exploring the Equations of
  Life} (Harvard University Press) 2006.

\bibitem{nowak1992evolutionary}
\Name{Nowak M.~A. \and May R.~M.} \REVIEW{Nature}{359}{1992}{826}.

\bibitem{nowak1993spatial}
\Name{Nowak M.~A. \and May R.~M.} \REVIEW{International Journal of bifurcation
  and chaos}{3}{1993}{35}.

\bibitem{lindgren1994evolutionary}
\Name{Lindgren K. \and Nordahl M.~G.} \REVIEW{Physica D}{75}{1994}{292}.

\bibitem{durrett1994importance}
\Name{Durrett R. \and Levin S.} \REVIEW{Theoretical Population
  Biology}{46}{1994}{363}.

\bibitem{killingback1996spatial}
\Name{Killingback T. \and Doebeli M.} \REVIEW{Proceedings of the Royal Society
  B-Biological Sciences}{263}{1996}{1135}.

\bibitem{szabo1998evolutionary}
\Name{Szab{\'o} G. \and T{\H{o}}ke C.} \REVIEW{Physical Review
  E}{58}{1998}{69}.

\bibitem{lieberman2005evolutionary}
\Name{Lieberman E., Hauert C. \and Nowak M.~A.}
  \REVIEW{Nature}{433}{2005}{312}.

\bibitem{du2008evolutionary}
\Name{Du W.-B., Zheng H.-R. \and Hu M.-B.} \REVIEW{Physica a-Statistical
  Mechanics and Its Applications}{387}{2008}{3796}.

\bibitem{allen2012mutation}
\Name{Allen B., Traulsen A., Tarnita C.~E. \and Nowak M.~A.} \REVIEW{Journal of
  Theoretical Biology}{299}{2012}{97}.

\bibitem{szabo2007evolutionary}
\Name{Szab{\'o} G. \and F{\'a}th G.} \REVIEW{Physics Reports-Review Section of
  Physics Letters}{446}{2007}{97}.

\bibitem{perc2010coevolutionary}
\Name{Perc M. \and Szolnoki A.} \REVIEW{Biosystems}{99}{2010}{109}.

\bibitem{hauert2004spatial}
\Name{Hauert C. \and Doebeli M.} \REVIEW{Nature}{428}{2004}{643}.

\bibitem{doebeli2005models}
\Name{Doebeli M. \and Hauert C.} \REVIEW{Ecology Letters}{8}{2005}{748}.

\bibitem{tomassini2006hawks}
\Name{Tomassini M., Luthi L. \and Giacobini M.} \REVIEW{Physical Review
  E}{73}{2006}{}.

\bibitem{ohtsuki2006simple}
\Name{Ohtsuki H., Hauert C., Lieberman E. \and Nowak M.~A.}
  \REVIEW{Nature}{441}{2006}{502}.

\bibitem{wu2007evolutionary}
\Name{Wu Z.-X., Guan J.-Y., Xu X.-J. \and Wang Y.-H.} \REVIEW{Physica
  a-Statistical Mechanics and Its Applications}{379}{2007}{672}.

\bibitem{abramson2001social}
\Name{Abramson G. \and Kuperman M.} \REVIEW{Physical Review E}{63}{2001}{}.

\bibitem{szolnoki2007cooperation}
\Name{Szolnoki A. \and Szab{\'o} G.} \REVIEW{Epl}{77}{2007}{}.

\bibitem{szolnoki2008towards}
\Name{Szolnoki A., Perc M. \and Danku Z.} \REVIEW{Physica a-Statistical
  Mechanics and Its Applications}{387}{2008}{2075}.

\bibitem{luthi2009evolutionary}
\Name{Luthi L., Tomassini M. \and Pestelacci E.}
  \REVIEW{Biosystems}{96}{2009}{213}.

\bibitem{ohtsuki2007evolutionary}
\Name{Ohtsuki H., Pacheco J.~M. \and Nowak M.~A.} \REVIEW{Journal of
  Theoretical Biology}{246}{2007}{681}.

\bibitem{ohtsuki2007breaking}
\Name{Ohtsuki H., Nowak M.~A. \and Pacheco J.~M.} \REVIEW{Physical Review
  Letters}{98}{2007}{}.

\bibitem{santos2005scale}
\Name{Santos F.~C. \and Pacheco J.~M.} \REVIEW{Physical Review
  Letters}{95}{2005}{}.

\bibitem{fu2007evolutionary}
\Name{Fu F., Liu L.~H. \and Wang L.} \REVIEW{European Physical Journal
  B}{56}{2007}{367}.

\bibitem{han2008evolutionary}
\Name{Yang H.-X., Gao K., Han X.-P. \and Wang B.-H.} \REVIEW{Chinese Physics
  B}{17}{2008}{2759}.

\bibitem{cassar2007coordination}
\Name{Cassar A.} \REVIEW{Games and Economic Behavior}{58}{2007}{209}.

\bibitem{grujic2010social}
\Name{Gruji{\'c} J., Fosco C., Araujo L., Cuesta J.~A. \and S{\'a}nchez {\'A}.}
  \REVIEW{Plos One}{5}{2010}{}.

\bibitem{kirchkamp2007naive}
\Name{Kirchkamp O. \and Nagel R.} \REVIEW{Games and Economic
  Behavior}{58}{2007}{269}.

\bibitem{gracia2012heterogeneous}
\Name{Gracia-L{\'a}zaro C., Ferrer A., Ruiz G., Taranc{\'o}n A., Cuesta J.~A.,
  S{\'a}nchez A. \and Moreno Y.} \REVIEW{Proceedings of the National Academy of
  Sciences}{109}{2012}{12922}.

\bibitem{hauert2002effects}
\Name{Hauert C.} \REVIEW{International Journal of Bifurcation and
  Chaos}{12}{2002}{1531}.

\bibitem{watts1998collective}
\Name{Watts D.~J. \and Strogatz S.~H.} \REVIEW{Nature}{393}{1998}{440}.

\bibitem{albert2002statistical}
\Name{Albert R. \and Barab{\'a}si A.~L.} \REVIEW{Reviews of Modern
  Physics}{74}{2002}{47}.

\bibitem{barabasi1999emergence}
\Name{Barabasi A.~L. \and Albert R.} \REVIEW{Science}{286}{1999}{509}.

\bibitem{traulsen2007pairwise}
\Name{Traulsen A., Pacheco J.~M. \and Nowak M.~A.} \REVIEW{Journal of
  Theoretical Biology}{246}{2007}{522}.

\bibitem{GoG:arXiv_Jinshan}
\Name{Zhang Q., Qi T., Li K., Di Z. \and Wu J.}
  \REVIEW{arXiv:1309.6715}{}{2013}{}.

\bibitem{chen2008promotion}
\Name{Chen X. \and Wang L.} \REVIEW{Physical Review E}{77}{2008}{}.

\bibitem{du2009evolutionary}
\Name{Du W.-B., Cao X.-B., Zhao L. \and Hu M.-B.} \REVIEW{Physica a-Statistical
  Mechanics and Its Applications}{388}{2009}{4509}.

\bibitem{chen2010evolutionary}
\Name{Chen C.-L., Cao X.-B., Du W.-B. \and Rong Z.-H.} \REVIEW{International
  Conference on Complexity and Interdisciplinary Sciences: 3rd China-Europe
  Summer School on Complexity Sciences}{3}{2010}{1845}.

\end{thebibliography}
\end{document}